\documentclass[longauth]{aa}  
\usepackage{graphicx}
\usepackage{txfonts}
\usepackage[colorlinks=true,linkcolor=blue,citecolor=blue]{hyperref}
\usepackage[toc,page]{appendix}
\usepackage{float}
\usepackage{subfloat}
\usepackage{subcaption}
\usepackage[flushleft]{threeparttable}

\newcommand{\feii}{\mbox{Fe\,{\sc ii}}}
\newcommand{\znii}{\mbox{Zn\,{\sc ii}}}
\newcommand{\crii}{\mbox{Cr\,{\sc ii}}}

\newcommand{\siii}{\mbox{Si\,{\sc ii}}}

\newcommand{\mgi}{\mbox{Mg\,{\sc i}}}

\newcommand{\hi}{\mbox{H\,{\sc i}}}

\newcommand{\niii}{\mbox{Ni\,{\sc ii}}}
\newcommand{\alii}{\mbox{Al\,{\sc ii}}}
\newcommand{\aliii}{\mbox{Al\,{\sc iii}}}

\newcommand{\civ}{\mbox{C\,{\sc iv}}}
\newcommand{\siiv}{\mbox{Si\,{\sc iv}}}

\begin{document}

\title{A large, chemically enriched, neutral gas reservoir\\in a galaxy at {\slshape z} = 6.782 \thanks{Based on observations carried out under ESO prog. ID 110.24CF.015 (PI: N. Tanvir) with the X-shooter spectrograph installed at the Cassegrain focus of the Very Large Telescope
(VLT), Unit 3 – Melipal, operated by the European Southern Observatory (ESO) on Cerro Paranal, Chile. }}

\subtitle{}

\author{A. Saccardi\inst{1}\fnmsep\thanks{E-mail: andrea.saccardi@cea.fr}, S. D. Vergani\inst{2,3}, L. Izzo\inst{4,5}, V. D'Elia\inst{6}, K. E. Heintz\inst{5,7,8}, A. De Cia\inst{9}, D. B. Malesani\inst{5,7,10}, J. T. Palmerio\inst{1}, P. Petitjean\inst{3}, S. Savaglio\inst{11,12,13}, N. R. Tanvir\inst{14}, R. Salvaterra\inst{15}, R. Brivio\inst{16}, S. Campana\inst{16}, L. Christensen\inst{5,7}, S. Covino\inst{16}, J. P. U. Fynbo\inst{5,7}, D. H. Hartmann\inst{17}, C. Konstantopoulou\inst{8}, A. J. Levan\inst{10,18}, A. Martin-Carrillo\inst{19}, A. Melandri\inst{20}, L. Piro\inst{21}, G. Pugliese\inst{22}, P. Schady\inst{23}, B. Schneider\inst{24}}

\institute{
Université Paris-Saclay, Université Paris Cité, CEA, CNRS, AIM, 91191, Gif-sur-Yvette, France
\and
LUX, Observatoire de Paris, Université PSL, CNRS, Sorbonne Université, 92190 Meudon, France
\and
Institut d’Astrophysique de Paris, UMR 7095, CNRS-SU, 98 bis
boulevard Arago, 75014, Paris, France
\and
INAF - Osservatorio Astronomico di Capodimonte, Salita Moiariello 16, 80131, Napoli, Italy
\and 
Niels Bohr Institute, University of Copenhagen, Jagtvej 128, 2200, Copenhagen, Denmark
\and
Space Science Data Center (SSDC) - Agenzia Spaziale Italiana
(ASI), I-00133 Roma, Italy
\and
Cosmic Dawn Center (DAWN), Denmark
\and
Department of Astronomy, University of Geneva, Chemin Pegasi 51, 1290 Versoix, Switzerland 
\and
European Southern Observatory, Karl-Schwarzschild Str. 2, 85748 Garching bei München, Germany
\and
Department of Astrophysics/IMAPP, Radboud University, 6525 AJ Nijmegen, The Netherlands
\and
Department of physics, University of Calabria, Via P. Bucci, Arcavacata di Rende (CS), Italy
\and
INAF – Osservatorio di Astrofisica e Scienza dello Spazio, Via Piero Gobetti 93/3, 40129 Bologna, Italy
\and
INFN – Laboratori Nazionali di Frascati, Frascati, Italy
\and
School of Physics and Astronomy, University of Leicester, University Road, Leicester LE1 7RH, UK
\and
INAF – IASF Milano, Via A. Corti 12, 20133 Milano, Italy
\and
Osservatorio Astronomico di Brera, via E. Bianchi 46, I23807, Merate (LC), Italy
\and
Department of Physics \& Astronomy, Clemson University, Clemson, SC 29634, USA
\and
University of Warwick, Coventry, CV4 7AL, UK
\and
School of Physics and Centre for Space Research, University College Dublin, Belfield D04 V1W8, Dublin, Ireland
\and
INAF - Osservatorio Astronomico di Roma, via Frascati 33, 00040
Monte Porzio Catone, Italy
\and
Istituto di Astrofisica e Planetologia Spaziali, Via Fosso del Cavaliere 100, 00133 Roma, Italy
\and
Astronomical Institute Anton Pannekoek, University of Amsterdam, 1090 GE Amsterdam, The Netherlands
\and
Department of Physics, University of Bath, Bath, BA2 7AY, UK
\and
Aix Marseille Universit´e, CNRS, CNES, LAM, Marseille, France
}

\date{}

 
  \abstract
   {The physical and chemical characterization of galaxies formed within the first billion years after the Big Bang remains one of the central goals in contemporary astrophysics. For the last two decades, optical and near-infrared spectroscopy of long gamma-ray bursts (GRBs) have been heralded as an effective diagnostic to probe the interstellar medium (ISM) of the galaxies hosting these events and their metal and dust content, reaching even the most distant redshifts.
   
   An opportunity to fulfill this expectation was provided by the recent blast triggered by the {\it Neil Gehrels Swift Observatory} of GRB\,240218A at redshift $z=6.782$. 
    From the GRB explosion, we could study a high-redshift galaxy selected in a complementary way with respect to flux-limited surveys, not depending on galaxy luminosity and stellar mass. 
   Furthermore, the GRB afterglow allows us to perform a coring of the galaxy ISM and to determine its kinematic and chemical properties. 
   We present the VLT/X-shooter spectrum of its afterglow enabling the detection and the detailed characterization of neutral-hydrogen, low-ionization, high-ionization and fine-structure absorption lines, as well as excited level transitions. From this rich variety of absorption lines associated with gas inside and around the GRB host galaxy, we determine the metallicity, kinematics, chemical abundance pattern and dust depletion. This provides the first detailed characterization of the neutral gas of a galaxy at $z>6.5$.

   Thanks to the presence of fine-structure absorption lines
   we could estimate the distance of the closest absorbing gas clouds as $d_{II}=620^{+230}_{-140}$ pc.
   We determine a high neutral hydrogen column density, $\log (N$(\hi{})/cm$^{-2})=22.5\pm0.3$, which is the highest one at 
   $z\gtrsim6$ determined so far for a GRB host galaxy, as well as a surprisingly high metal column density, $\log (N$(\znii)/cm$^{-2})>14.3$. 
   The observed metallicity of the host galaxy system is $\mathrm{[Zn/H]>-0.8}$.
   Taking the derived column densities at face value, although a number of transitions are saturated, we find evidence of a high amount of dust depletion and of peculiar nucleosynthesis, with an overabundance of aluminum, as in other high-redshift GRB host galaxies at $z\sim6$.

   The high hydrogen column density, metal abundances and dust depletion in the neutral gas align with those of the ionized gas of very high-redshift galaxies unveiled by ALMA and JWST, testifying that a rapid build up of metals and dust, and massive neutral hydrogen reservoirs seem to be common features of galaxies in the early Universe. This research highlights the remarkable potential of GRBs as tools to investigate the detailed properties of galaxies deep into the reionization epoch, and stresses the importance of new missions  capable to enlarge the sample of very high-redshift GRBs.

  }


   \titlerunning{A large, chemically enriched, neutral gas reservoir in a galaxy at {\slshape z} = 6.782}
   \authorrunning{A. Saccardi et al.}

   \maketitle
%
\section{Introduction}
Unveiling galaxies at the highest redshifts and studying their chemical evolution is a key objective in modern astrophysics.

Neutral gas, primarily composed of atomic hydrogen (\hi{}), is a fundamental component of the interstellar medium (ISM) and the circumgalactic medium (CGM) in galaxies, and contains the majority of metals at $z\geq2.5$ \citep{Peroux2020}. 
Neutral gas forms the primary reservoir from which stars form and hence a key player in chemical enrichment.
Therefore, its observation is fundamental to understand galaxy evolution \citep{Madau2014}.

Nowadays, thanks to {\it{James Webb}} Space Telescope (JWST), it is possible to measure \hi\,column densities in large fractions in galaxies during the reionization epoch at $z>5$ and out to $z\sim 14$ \citep{Umeda2024,D'Eugenio2024,Hainline2024,Carniani2024_dla,Witstok2024_dla,Heintz2025}. 
Through the detection of strong damped Lyman-$\alpha$ (DLA) absorbers, \cite{Heintz2024} provided direct measurements of \hi\,for galaxies at $z>9$, and showed evidence for abundant neutral gas reservoirs in some early galaxies. In order to sustain the high \hi\,column densities while covering the entire physical extent of the galaxies, these sources must be embedded in large, extended layers or shells of neutral gas. High-redshift galaxies at $z>6$ were not only forming stars efficiently but were also undergoing significant chemical evolution \citep{Langeroodi2023,Heintz2023Nat_jwst,Nakajima2023,Curti2024}, showing richness of spectral features and unusual chemical abundances \citep[e.g.][]{Bunker2023,Curti2024_galaxy9.4,Castellano2024,D'Eugenio2024}, which may indicate rapid enrichment processes occurring during the earliest stages of galaxy formation. These findings challenge earlier models that suggested that metal enrichment and dust formation were slower processes that occurred after $z\sim5$ \citep{Madau2014}.

Despite this progress, a detailed study of chemical and kinematic properties of neutral gas remains difficult, even for JWST. The most powerful way to directly measure the properties of the neutral gas is through absorption lines detected in the spectrum of a bright background source, such as quasars (QSOs) and gamma-ray bursts \citep[GRBs;][]{Gehrels2013}. GRBs serve as cosmic beacons, that probe in great detail the star-forming galaxies that host them, which are often faint and missed by flux-limited observations, out to the highest redshifts. The association of long-GRBs (LGRBs) with massive stars \citep{Hjorth2003,Woosley2006a,Fruchter2006,Hjorth2012,Cano2017} means that they point directly to the sites of star formation enriching the Universe \citep{Krogager2024}, unlike the alignments of QSOs, and makes them especially suitable to investigate galaxies and star formation up to the early Universe. They have already been observed out to $z_{spec}=8.23$ \citep[GRB 090423A;][]{Salvaterra2009,Tanvir2009} and $z_{phot}\simeq9.4$ \citep[GRB 090429B;][]{Cucchiara2011}, and they are expected to be observable even beyond \citep{Campana2022,Kann2024}.

Ground-based telescopes equipped with medium/high-resolution spectrographs allow us to dissect the light from GRB afterglows, revealing precise information about the chemical and physical properties of the absorbing gas in the host galaxy, in particular on the neutral gas and its components. Optical and near-infrared (NIR) absorption spectroscopy and the high spectral signal-to-noise ratio (SNR) provide valuable information on the neutral hydrogen column density \citep{Selsing2019,Tanvir2019}, the gas kinematics, and the distance of absorbing material from the GRB \citep{Prochaska2006,Vreeswijk2007,DElia2009}, shedding light on the chemical properties of the ISM, metals, and dust content of their host galaxies up to the highest redshifts \citep{Sparre2014,Hartoog2015,Saccardi23}. All of this can be done independently of the brightness of the host galaxies. Furthermore, thanks to Very Large Telescope (VLT) X-shooter follow-up and its much higher spectral resolving power, we can constrain the neutral gas-phase and chemical composition much more accurately than will ever be possible with JWST.

Several studies investigate the metallicity and dust content of GRB-selected star-forming galaxies from $z\sim2$ up to $z>6$ \citep[e.g.,][]{Bolmer2019,Heintz2023}. These galaxies exhibit significant metal and dust production, highlighting a rapid build-up of these components in the early Universe. They also show a large scatter in the dust-corrected metallicities at a given redshift \citep{DeCia2018,Heintz2023,Konstantopoulou2024}, which is not captured in most state-of-the-art galaxy evolution simulations \citep{Yates2021}, although there is broad agreement with the chemical enrichment as a function of cosmic time. Furthermore, these works reveal that the dust-to-metal ratio in these galaxies grows steadily with cosmic time, aligning with the predictions of earlier models, but suggesting faster dust production than previously assumed. Additionally, there is evidence for a correlation between metallicity and dust-to-metal ratio \citep{DeCia2013}, suggesting that these galaxies are significant contributors to the cosmic dust budget at high redshifts \citep[see][for further details]{Heintz2023}.

All the results presented above exploited the best existing samples of GRB hosts, but are strongly impacted by poor statistics due to the small number of events, especially at high-redshift. To date, GRB\,130606A \citep{Hartoog2013} and GRB\,210905A \citep{Saccardi23}, at $z=5.913$ and $z=6.312$, respectively, are the highest redshift GRBs with available optical and NIR spectroscopy of their afterglow characterized by high SNR, good spectral resolution and a large number of detected absorption lines of the neutral gas in their host galaxies. With these afterglow spectra it was possible to study the chemical abundance pattern and depletion of metals, and to retrieve information on nucleosynthesis, all in an unprecedented detail for faint galaxies at such high-redshift \citep{Chornock2013,Totani2014,Hartoog2015,Rossi2022,Saccardi23,Fausey2025}.

In this paper, we study the VLT/X-shooter optical and NIR afterglow spectrum of GRB\,240218A at redshift $z=6.782$. In Sect. \S\ref{data} and \S\ref{data analysis}, we present our dataset, the fitting of the absorption lines, together with the curve of growth analysis and the GRB-absorbers distance calculation. In Sect. \S\ref{results} we present our results. Discussion and conclusions are drawn in Sect. \S\ref{conclusions}. A $\Lambda$CDM cosmological model with $\Omega_{M}=0.308$, $\Omega_{\Lambda}=0.692$, and $H_{0}=67.8$\,km s$^{-1}$ Mpc$^{-1}$ \citep{Planck2016} has been assumed for calculations. All data are in observer frame and 1\,$\sigma$ errors are reported throughout the paper, unless otherwise specified.
A companion paper \citep{Brivio2025} presents and analyzes the GRB prompt and afterglow multiwavelength spectral and temporal properties, comparing with those of the few high-redshift LGRBs observed so far and those of lower-$z$ events.


\section{VLT/X-shooter observation of GRB 240218A}
\label{data}

On February 18, 2024, at 02:00:00 UT the \textit{Neil Gehrels Swift Observatory} (\textit{Swift} hereafter; \citealt{Gehrels2004}) discovered GRB\,240218A. The \textit{Swift} Burst Alert Telescope (BAT, \citealt{Barthelmy2005}) triggered and located the explosion \citep{gcnswift}. \textit{Swift} slewed immediately to the burst and the X-Ray Telescope (XRT, \citealt{Burrows2005}) began observing the field 149.4 seconds after the BAT trigger. XRT found a bright, uncatalogued X-ray source at the enhanced position of coordinates RA (J2000) = 10h 47m 11.24s Dec. (J2000) = $+$01\degr{} 16$'$ 34.8\arcsec{} with an uncertainty of 4.2\arcsec{} \citep{gcnswiftenhanced}. The {\it Swift} Ultra-Violet and Optical Telescope (UVOT, \citealt{Roming2005}) took a finding chart exposure of 119 seconds with the White filter  starting 158~s after the BAT trigger. No credible afterglow candidate was found in the data products. Follow-up from the ground allowed the detection of a bright NIR afterglow candidate \citep{gcnrem,lbt_gcn_nir} and its optical counterpart \citep{fors_gcn_z}, together with a precise localization of the afterglow thanks to the radio \citep{gcnvla,atca_gcn_radio} and mm \citep{gcnalma} detections (see \citealt{Brivio2025} for further details).

\begin{table}[h]
\centering
\tiny
\caption{Log of the observations.}
 \begin{tabular}{lccccc}
  \hline
  \hline
  Epoch & Arm & Exp. time & Wav. range & Slit & Resolution $R$\\
  (Hours) &  & (s) & $(\AA)$ & (${\prime \prime}$) & $(\lambda/\delta\lambda)$\\ 
  \hline
26.47 & UVB & 4x1200 & 3000-5600 & 1.0x11 & 5400\\
26.47 & VIS & 4x1200 & 5600-10200 & 0.9x11 & 8900\\
26.47 & NIR & 8x600 & 10200-21000 & 0.9x11JH & 5600\\
  \hline
  \hline
 \end{tabular}
  \label{tab:1}
\end{table}

Approximately $\sim26$~hr (observer frame) after the detection of GRB\,240218A, we observed its afterglow using the ESO VLT UT3 equipped with the X-shooter spectrograph \citep{Vernet2011}. The observing setup is detailed in Table \ref{tab:1}. Observations were conducted using the ABBA nod-on-slit mode. Each individual spectrum from the UVB and VIS arms was reduced using the STARE mode reduction, with the extraction window positioned at the GRB afterglow trace. The NIR arm data were processed using the standard X-shooter NOD mode pipeline \citep{Goldoni2006, Modigliani2010}. Sky features were subtracted and each flux-calibrated spectrum was combined into a final science spectrum. Additionally, a telluric correction was applied to the final stacked VIS and NIR spectra, and wavelengths were corrected to the vacuum-heliocentric system. To ensure a correct flux calibration between the VIS and NIR arms, both spectra were normalized to the afterglow photometry in the $z$ and $J$ band, respectively, reported at the mean time of the spectrum using the light curve fit provided by \cite{Brivio2025} which presents a complete analysis of the multiwavelength light curve. An integration of the spectrum over the filter profile was necessary given that the \hi{} absorption (Lyman-$\alpha$ break) significantly affects the $z$-band magnitude.


\section{Data analysis}
\label{data analysis}

The highest redshift system identified in the absorption spectrum of GRB\,240218A afterglow is at $z=6.782$ and spans $\sim$$400$\,km\,s$^{-1}$. It is a DLA absorption system (see Figure \ref{nh}) with associated metal absorption lines \citep{gcnvlt}. As this is the highest redshift \hi{} absorption in the spectrum and, furthermore, fine-structure absorption lines are detected corresponding to this system, we associate it with the host galaxy of GRB\,240218A. We do not identify any foreground intervening absorbers along the line of sight.
We note that this is the highest GRB redshift measured accurately from metal absorption lines so far.

\subsection{The damped Ly-$\alpha$ absorption feature}
\begin{figure}[h]
    \centering
    \includegraphics[scale=0.7]{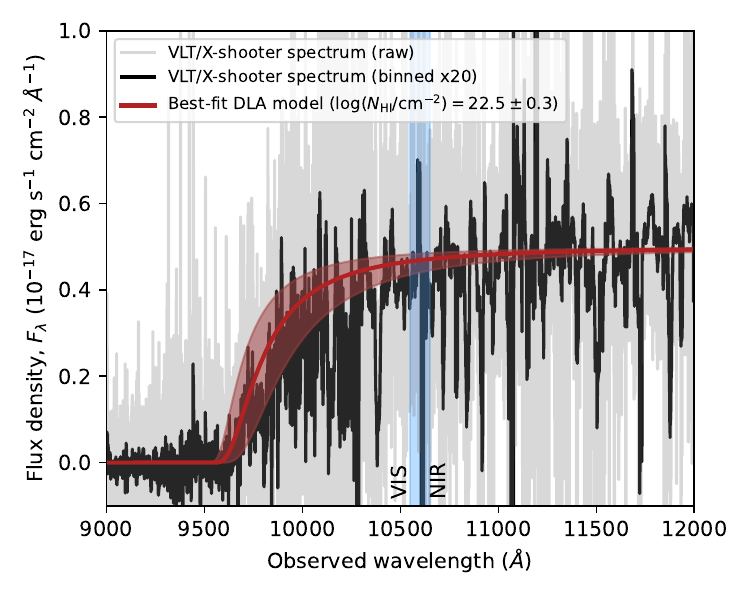}
    \caption{VLT/X-shooter 1D spectrum of GRB\,240218A. The grey curve show the raw, stitched VIS+NIR arm, photometrically calibrated spectrum, and the black the binned version (by a factor of 20). The best-fit DLA model with $\log (N_{\rm HI}/{\rm cm}^{-2}) = 22.5\pm 0.3$ is shown as the red curve, with the uncertainty represented by the red-shaded area. The connection region between the VIS and NIR arms is represented by the blue-shaded area.}
    \label{nh}
\end{figure}

The DLA absorption imprinted from \hi{} in the host-galaxy ISM is shown in Figure \ref{nh}.

The determination of the DLA column density is complicated by the fact that the red wing falls in a noisy region of the spectrum at the overlap of the VIS and NIR arms. At this high redshift, we also expect a significant neutral \hi{} fraction in the line of sight through the partially neutral IGM.

We model the Ly$\alpha$ damping wing with a Voigt profile, with an optical depth $\tau = C a H(a,x) N_{\rm HI}$, where $C$ is the photon absorption constant, $a$ is the damping parameter, and $H(a, x)$ is the Voigt-Hjerting function, following the approximation of \cite{TepperGarcia06}. We further add the optical depth from the Gunn-Peterson effect from an increasingly neutral intergalactic medium (IGM; \citealt{Fan2006}), following the approximation by \citet{MiraldaEscude1998,Totani2006}, integrating the line-of-sight IGM contribution from the GRB redshift down to $z=6$. We assume that the intrinsic slope can be quantified as a power-law, $F_\lambda \propto \lambda^{-\beta}$, and simultaneously constrain the three free parameters in the fit, $\beta$, $N_{\rm HI}$, and $x_{\rm HI}$. Since the neutral \hi{} fraction of the IGM, $x_{\rm HI}$, is essentially unconstrained in the modeling due to the substantial local \hi{} component in the GRB host, we fix this quantity to $x_{\rm HI}=0.2$ in the final best-fit model. Assuming a single velocity component fixed to the systemic redshift $z_{\rm GRB} = 6.782$, a final value of $\log (N$(\hi{})/cm$^{-2})=22.5\pm0.3$ is obtained.

\begin{figure*}
    \centering
    \includegraphics[scale=1]{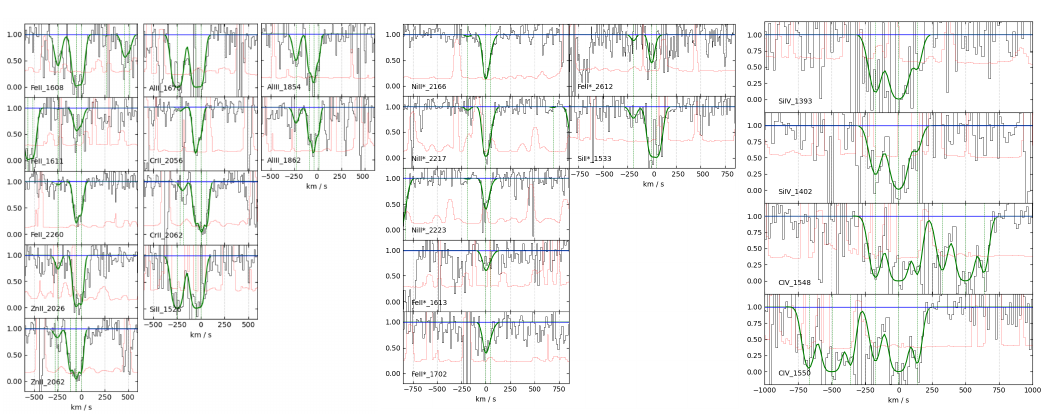}
    \caption{VLT/X-shooter optical/NIR afterglow spectrum of GRB\,240218A. {\it Left panel}: selection of low-ionization absorption lines of the GRB host galaxy system. Here and in the following panels data are in black, the fit is in green, the error spectrum is in red, the continuum in blue and the vertical green dashed lines indicate the center of the components. {\it Middle panel}: fine-structure and excited transition absorption lines of the GRB host galaxy system. {\it Right panel}: high-ionization absorption lines. All the plots are in velocity space and 0 has been fixed to $z=6.782$ (see Sect. \ref{data analysis}), corresponding to the stronger low-ionization line component (II).}
    \label{abs_spec}
\end{figure*}

\begin{table*}
\centering
\caption{{{\it Top}}: Column density of low-ionization lines. {{\it Bottom}}: Column density of high-ionization lines. The last row reports the Doppler parameter $b$ used of each component as resolved by the X-shooter observations. Results assuming column densities as measurements are reported in square brackets.}
\small
\centering
\begin{tabular}{c c c c}       
\hline\hline
Species & $I$ & $II$ & $III$ \\
\hline
& $-200\,{\rm km~s}^{-1}$ & $0\,{\rm km~s}^{-1}$ & $+50\,{\rm km~s}^{-1}$ \\
\hline

\alii{} $\lambda1670$ & $>14.7$ & $>15.0$ & $>14.5$ \\

\znii{} $\lambda2026,\lambda2062$ & $12.9\pm0.2$ & $>14.1\,[14.1\pm0.2]$ & $>13.9\,[13.9\pm0.1]$\\

\crii{} $\lambda2056,\lambda2062$ & $13.0\pm0.2$ & $>14.4\,[14.4\pm0.1]$ & $>13.8\,[13.8\pm0.1]$\\

\siii{} $\lambda1526$ & $>15.9$ & $>16.5$ & $>15.5$\\

\feii{} $\lambda1608, \lambda1611, \lambda2260$$^\ddagger$ & $14.6 \pm 0.1$ & $>15.8\,[15.8\pm 0.1]$ & $>15.6\,[15.6\pm 0.1]$\\

\feii{}{*} $\lambda2612$ (1s) & $<13.3$ & $13.8 \pm 0.1$ & $12.6 \pm 0.4$\\

\feii{}{*} $\lambda1613, \lambda1702$ (5s) &  & $14.5 \pm 0.1$ & $13.8 \pm 0.3$\\

\siii{}{*} $\lambda1533$ & $<12.4$ & $>14.8$ & $>15.1$\\

\niii{}{*} $\lambda2166,\lambda2217,\lambda2223$ & $<12.9$ & $14.2 \pm 0.1$ & $13.3 \pm 0.1$\\

\aliii{} $\lambda$1854, $\lambda$1862$^\dagger$  & $13.3\pm0.2$ & $>14.1$ & $>13.2$  \\

\hline
\hline
$b\,{\rm (km~s^{-1})}^*$ & $34\pm4$ & $27\pm4$ & $21\pm4$ \\
\hline
\hline
\end{tabular}

\vspace{0.5cm}

\begin{tabular}{c c c c } 
\hline\hline
Species & $1$ & $2$ & $3$ \\
\hline
& $-200\,{\rm km~s}^{-1}$ & $0\,{\rm km~s}^{-1}$& $+140\,{\rm km~s}^{-1}$\\
\hline
\civ{}$\lambda1548$, $\lambda1550$$^\mathsection$ & $>14.7$ & $>15.7$ & $>15.8$ \\
\siiv{}$\lambda1393$, $\lambda1402$$^\mathsection$ & $>14.3$ & $>15.2$ & $>13.7$\\
\hline        
\hline
$b\,{\rm (km~s^{-1})}$ & $38\pm4$ & $44\pm4$ & $17\pm4$\\
\hline
\hline

\end{tabular}

\begin{tablenotes}
    \item$^\ddagger$ Other \feii{} transitions are identified in the absorption spectrum e.g. \feii{} $\lambda2344, \lambda2374, \lambda2587, \lambda2600$. Unfortunately, they fall in a region of the spectrum strongly affected by telluric absorption and with poor SNR. For this reason they are not taken into account when performing the Voigt-profile fitting. \item $\dagger$ Component II of \aliii{} is fitted with a different Doppler parameter, $b\,{\rm (km~s^{-1})}=47\pm4$, due to its different ionization energy.
    \item $^\mathsection$ \civ{} and \siiv{} lines are particularly uncertain because they are found in a very noisy region of the spectrum.
    \item $^*$ The determined b-values are rather high and this probably means that the components are composed of multiple narrower ones.
\end{tablenotes}
\label{table_N} 
\end{table*}

\subsection{Metal absorption lines}

The low-ions (\feii{}, \crii{}, \znii{}, \siii{}, \alii{}, etc.) trace the neutral medium in the host galaxy.
In GRB\,240218A afterglow spectra, low-ionization lines are detected in three separate components (see Figure \ref{abs_spec} and Table \ref{table_N}). Component II is the strongest and shows fine-structure absorption lines; we therefore adopt its redshift ($z_{\rm GRB} = 6.782$) as that of the GRB host (refer also to analysis in Sect. \ref{fine-stru-analysis} on distance of absorbing gas clouds) and as the zero reference value in velocity space. Components I and III are at $\Delta$v$\,= -200$\,km\,s$^{-1}$ and $\Delta$v$\,= +50$\,km\,s$^{-1}$, respectively.

For components II and III, the spectrum shows clear detection of transitions arising from excited levels of \feii{}, and from metastable levels (involving transitions from long-lived excited states that are forbidden or suppressed by selection rules) of both \feii{} and \niii{}. The \feii{}$^*\,\lambda2612$, \feii{}$^*\,\lambda1613$, \feii{}$^*\,\lambda1702$ and \niii{}$^*\,\lambda2166$, \niii{}$^*\,\lambda2217$, \niii{}$^*\,\lambda2223$ transitions are identified as shown in Figure \ref{abs_spec}. A comprehensive analysis of fine-structure absorption lines and cloud distances is presented in Sect. \ref{fine-stru-analysis} and Appendix \ref{app_a}.

Despite the very noisy region of the spectrum, high-ionization lines (\civ{}, \siiv{}) are also detected, with their strongest component aligned with that of low-ionization transitions. They span a larger range with respect to low-ionization lines in velocity space, $\Delta$v$\, \sim 500$\,km\,s$^{-1}$ (see Figure \ref{abs_spec}).

\begin{figure}[!htbp]
    \centering
    \includegraphics[scale=0.42]{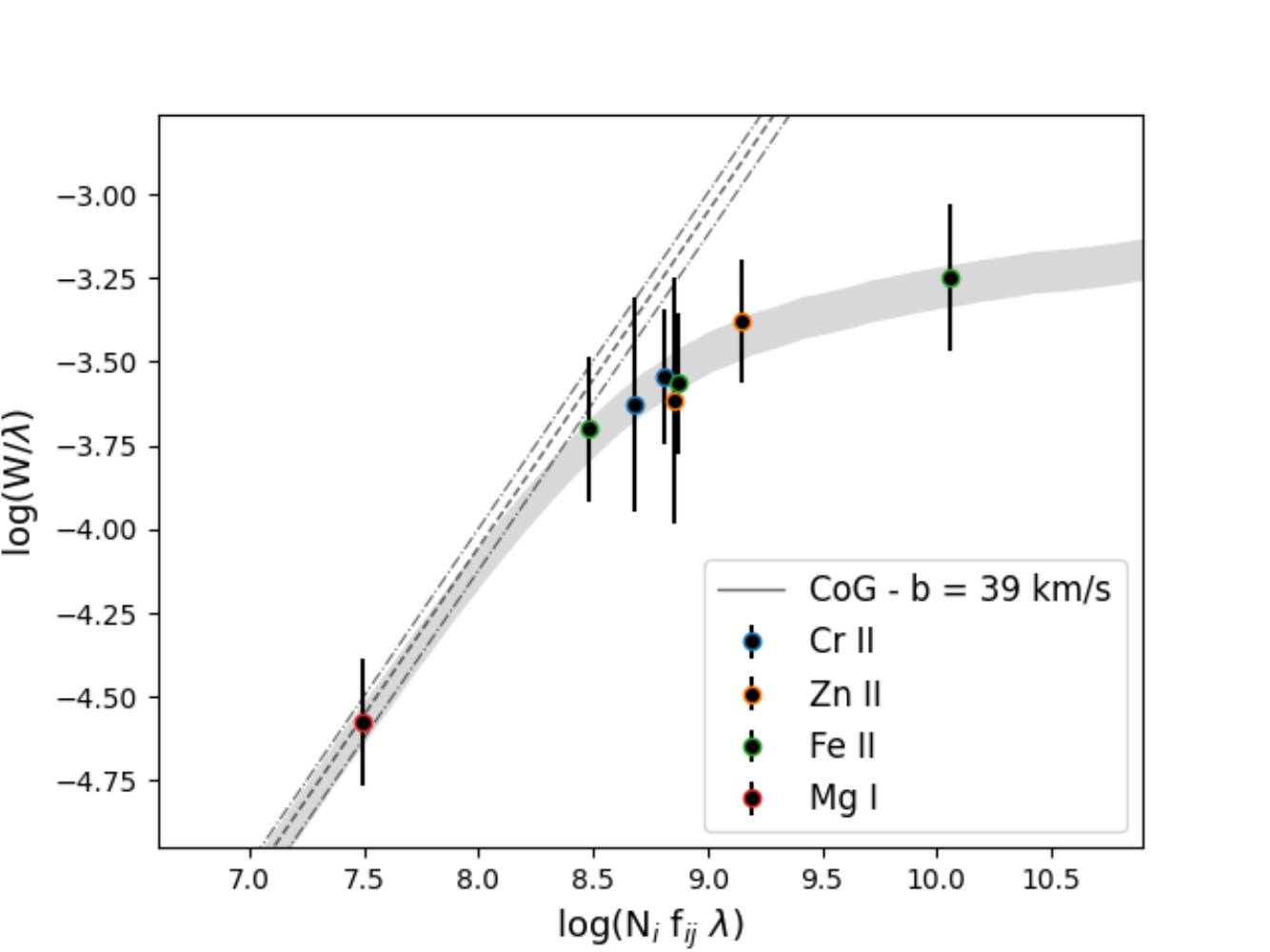}
    \caption{Results obtained from the CoG analysis on the ISM absorption lines identified in the spectrum of GRB\,240218A. The data points are color-coded with respect to different element transitions (see Table \ref{tab:EW}). The grey shaded area represents the best fit model which results in the labeled value of the Doppler parameter, while the dotted lines represent the linear approximation regime and its 1$\sigma$ confidence ranges.}
    \label{cog}
\end{figure}

Most of the low-ionization absorption lines show signs of saturation. The fitting procedure in the saturated regime, flat region of the Curve of Growth (CoG), is not straightforward. Fitting simultaneously multiple transitions of the same ionic element can address the hidden saturation issue. An intermediate-resolution spectrograph, such as X-shooter, can reduce this effect slightly, but caution is needed when interpreting column densities obtained via CoG. Such measurements should generally be considered as lower limits, especially if the equivalent widths ($EW$s) fall within the saturated region of the CoG \citep{Prochaska2006_cog}.
Indeed, at the X-shooter resolution, to get more reliable constraints one should use a combination of the lines with different oscillator strengths. To estimate the column densities of elements with similar ionization potential, we first use the CoG approach, as described in \citet{Spitzer1998} and explained in detail in Appendix \ref{app_b}. The best-fit CoG is illustrated in Figure \ref{cog}.
The results show a quite high  Doppler parameter $b$, likely indicating spectrally unresolved components, and that most of the transitions fall at the knee of the CoG (see Figure \ref{cog}), which is the transition region between linear and saturated regime.

It is evident that the large EW (see Table \ref{tab:EW}) of the metal lines (e.g. \feii{}\,$\lambda2260$, \znii{} doublet) is remarkable at such high redshift. They are not only significantly stronger than what is typically observed in GRB-DLAs \citep{deUgartePostigo2012}, but also comparable to the values commonly observed in the Milky Way \citep[MW; e.g.,][]{DeCia2021}.

We then performed a Voigt fit profile analysis to determine the element column densities. This is another approach which follows the same physical method as CoG with the only difference being that the latter, relying on EW measurements, does not depend on spectral resolution. We fit the systems with the {\it Astrocook} code \citep{Cupani2020}, a {\it Python} software environment to analyze absorption spectra that includes a set of algorithms to model spectral features in absorption and emission (continuum, spectral lines, complex absorption systems). The column densities that we determined are presented in Table \ref{table_N}. The errors reported are those obtained by the line fitting.

The simultaneous fit of all the transitions of the \crii{} multiplet\footnote{
The \crii{}\,$\lambda2066$ transition, which would be the weakest of the triplet (\crii{}\,$\lambda2056$, \crii{}\,$\lambda2062$, \crii{}\,$\lambda2066$) is strongly affected by a sky line.}(\crii{}\,$\lambda2056$, \crii{}\,$\lambda2062$) makes it possible to estimate the chromium column density. The flux residual of \crii{}\,$\lambda2056$ suggests that the \crii{} column density may be considered as a lower limit. 
Concerning iron, the \feii{}\,$\lambda2260$ transition shows as well flux residuals which are at the limit of saturation. The \feii{}\,$\lambda1611$ transition does not appear to be saturated as shown in Figure \ref{abs_spec}. It would confirm that the \feii{}\,$\lambda2260$ transition is not affected by severe saturation. However, given the bad SNR region where the $\lambda1611$ transition falls, we may also consider the \feii{} column density as a lower limit.
On the other hand, Zn is a more delicate case. Figure \ref{abs_spec} shows evidence for some saturation of \znii{}\,$\lambda2026$, while the \znii{}\,$\lambda2062$ is blended with \crii{}\,$\lambda2062$. By measuring the \crii{} column density with the Voigt-profile fitting, it is possible 
to deblend the lines and recover the contribution of \znii{}\footnote{The same method has been applied for \znii{}\,$\lambda2026$ which is blended with \mgi{}\,$\lambda2026$.}. As for the other elements it would be cautious to take the \znii{} column density as a lower limit to avoid incorrect conclusion and a systematic underestimate of all column density. We note that, with respect to our scientific results (see Sect. \ref{metallicity_subsec}), considering the \znii{} column density as a measurement is a conservative approach.

As expected, the Voigt fit approach and the CoG provide consistent results. Indeed, considering a single component even for the Voigt fit profile analysis, the Doppler parameter is in agreement with that determined by the CoG.

\section{Results}
\label{results}

\subsection{Fine-structure lines and absorbing cloud distances}
\label{fine-stru-analysis}

GRB afterglows strongly interact with the interstellar medium, by depositing a large amount of energy. This is particularly evident in the optical-UV band where the gas surrounding the GRB absorbs the radiation, exciting the atoms and ions to high quantum levels. The comparison of observations with predictions from time dependent photo-excitation codes has been applied to the spectra of several GRBs \citep{Dessauges-Zavadsky2006,DElia2009_variab, Ledoux2009,Hartoog2013,DElia2014,Saccardi23,Pugliese2024}, allowing estimates of the distance between the GRB and the absorber, which typically spans tens to hundreds of pc \citep{Vreeswijk2012}.

For GRB\,240218A we can apply the comparison between data and photo-excitation codes to find the distance between the GRB and the absorbers of components II and III (we refer to the Appendix \ref{app_a} for more details). We find $d_{II}=620^{+230}_{-140}$ pc, $d_{III}=2.0^{+7.6}_{-0.6}$ kpc, for component II and III, respectively. Despite the large uncertainties for component III due to the larger errors on the column densities, we can safely conclude that component II lies closer to the GRB than component III. 
We note that these distances could be larger if the ground state column densities are underestimated (if affected by saturation).

\subsection{Metallicity and chemical enrichment of the neutral gas}
\label{metallicity_subsec}

We used the metal and neutral hydrogen column densities to determine the metallicity of GRB\,240218A host galaxy along the GRB line of sight. Based on the column density of Zinc ($\log (N$(\znii)/cm$^{-2})>14.3$), that is a poorly dust depleted element, we calculate the observed metallicity $\mathrm{[X/H]>-0.8}$ (where X is Zn).

Despite being aware of the limitations of the spectral data, especially regarding the 
saturation issues, we report here a deeper analysis on the chemical enrichment and dust depletion by assuming column densities of \feii{}, \crii{}, and \znii{} as measurements. We used the metal and neutral hydrogen column densities reported in Table \ref{metallicity}, assuming \cite{Asplund2021} solar abundances.

\begin{table}[h!]     
\centering
\caption{Column densities, metal abundances, and relative abundances with respect to iron. Results assuming column densities as measurements are reported in square brackets.} 
\begin{tabular}{c c c c}       
\hline\hline
X & $\log(N/{\rm cm}{^{-2}})$ & [X/H] & [X/Fe] \\
\hline
Al & $>15.3$ & $>-1.7$ & $[>0.3]$\\
Zn & $>14.3\,[14.3\pm 0.2]$ & $[-0.8 \pm 0.4]$ & $[1.2\pm0.2]$\\
Si & $>16.6$ & $>-1.4$ & $[>0.6]$\\
Fe & $>16.0\,[16.0\pm 0.1]$ & $[-1.9\pm0.3]$ & \\
Cr & $>14.5\,[14.5\pm 0.1]$ & $[-1.6 \pm0.3]$ & $[0.3\pm0.1]$\\
\hline
\hline
\end{tabular}
\label{metallicity} 
\end{table}

Following the method developed by \citet{DeCia2016}, \citet{DeCia2021} (but using the most recent and solid compilation of the so-called refractory indices from \citealt{Konstantopoulou2024}), we analyze the abundances of different metals with the aim of characterizing the chemical enrichment in the GRB host galaxy. We stress that this deeper analysis on the chemical enrichment and dust depletion is based on the assumption of considering, despite the possible saturation issue, the column densities values as measures and not as limits.

\begin{figure}[h!]
    \centering
    \includegraphics[scale=0.6]{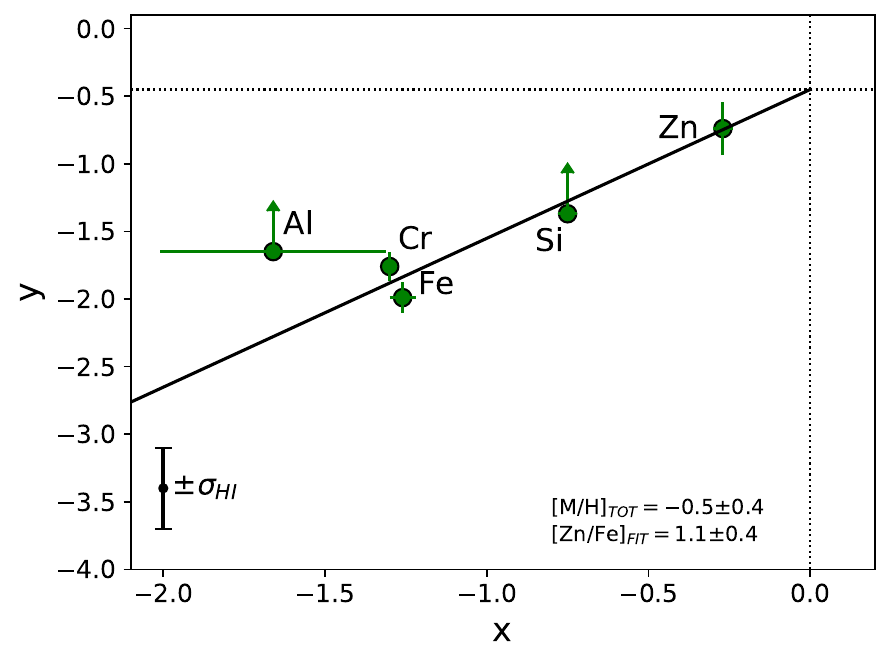}
    \caption{Abundance pattern observed in the host of GRB\,240218A from the total absorption-line profile. The observed abundances are on the $y$-axis, and the $x$-axis represent the tendency of each element to deplete into dust grains. The linear fit to the data (solid line) determines the dust depletion corrected metallicity [M/H]$_{\rm tot}$ (intercept) and the overall strength of depletion [Zn/Fe]$_{\rm fit}$ (slope). The \hi{} error bar, the same for all the data points, is reported as $\sigma_{HI}$. See \citet{DeCia2021,DeCia2024} and \citet{Konstantopoulou2024} for a exhaustive description of the method.}
    \label{dust_240218A_measure}
\end{figure}

Despite the large uncertainties and the limited amount of elements available
, the abundance pattern shown in Figure \ref{dust_240218A_measure} indicates clearly significant dust depletion. We fit a linear relation\footnote{For more details on the method please refer to \cite{DeCia2024} (i.e. in particular to Figure 3) and \cite{Konstantopoulou2024}; see also \cite{Saccardi23} for a direct application of such a method to the high-redshift case of GRB\,210905A.} to the constrained data (Cr, Fe, and Zn), and not including Al and Si\footnote{In previous high redshift GRB-DLAs there is the evidence of an Al overabundance (see Sect. \ref{results}). Its deviation from the linear fit is much more related to nucleosythesis effect rather than dust depletion. Hence, its inclusion in the fit could bias the results. We also exclude Si from the linear fit for the same reasons (nucleosynthesis). We thus consider them as lower limits and not taking into account during the fit procedure.}.

The dust-corrected metallicity is $\mathrm{[M/H]_{tot}=-0.5\pm0.4}$ with a dust depletion of $\mathrm{[Zn/Fe]_{fit}=1.1\pm0.4}$.
The large uncertainties are mainly due to the low SNR of the X-shooter spectrum. We note that considering \znii{} column density as a lower limit would imply even higher metallicity and dust depletion values.

Following \cite{Konstantopoulou2024} we derive the dust-to-metal mass ratio (DTM), that is the ratio between the mass of dust and the total mass of the metals. We find $DTM=0.5\pm0.1$. This DTM value, together with that of $\mathrm{[Zn/Fe]_{fit}}$, is similar to what is observed in the MW. The GRB-selected galaxies further show systematically lower DTM than the MW, especially at high-redshift \citep{Heintz2023}. Hence, GRB\,240218A DTM value represents an exception with respect to the best-fit relation derived from observed GRB-DLA \citep{Heintz2023} and values predicted by the simulations \citep{Li2019}.

The above results bring to a high value of $A_{\rm V,depl}\sim5$. The uncertainty on the extinction value derived from depletion ($A_{\rm V,depl}$) is also high due to the large uncertainties on the total dust-corrected metallicity and \hi\,column density, and it is not possible to place an accurate constraint on this measure. The amount of extinction needed to reproduce the afterglow spectral energy distribution (SED) is $A_{\rm V,SED}=0.35\pm0.03$~mag \citep{Brivio2025}.

Notably, we observe a sign of aluminium overabundance (see Figure \ref{dust_240218A_measure}, and Appendix \ref{app_d}). This is also the case in the other two high-$z$ GRB afterglows with a comprehensive analysis of the abundance patterns, GRB\,130606A \citep{Hartoog2015} at $z=5.913$ and GRB\,210905A \citep{Saccardi23} at $z=6.312$. This peculiar chemical pattern, nowadays found typically in globular clusters, can be the direct signature of the presence of very massive rotating stars \citep[][and references therein]{Prantzos2007}. Furthermore, we perform a component-by-component analysis of the depletion patterns. All show a strong amount of dust depletion, 
and component $I$ shows a clear evidence of aluminum overabundance. The details of this analysis are shown in Appendix \ref{app_d}. In Figures \ref{fit_comp} and \ref{abund_comp} we show the linear fit for each component and the over- and under-abundance of the different elements with respect to iron after correcting for dust depletion, respectively.

\section{Discussion and conclusions}
\label{conclusions}

We analyzed the gas properties along the GRB\,240218A line of sight using afterglow spectra obtained with VLT/X-shooter. Our study reveals the presence of neutral hydrogen, low-ionization, high-ionization, and fine-structure metal lines, originating from the GRB host galaxy complex at $z=6.782$. We measured a high neutral hydrogen column density in the host galaxy of $\log (N$(\hi{})/cm$^{-2})=22.5\pm0.3$.
We determined the abundances of metals in the neutral gas of the ISM along the GRB line of sight ($\mathrm{[Zn/H]>-0.8}$). Under the assumption that the column densities can be derived although aware of the saturation issue, we found evidences of high dust depletion, that we estimate to correspond to $\mathrm{[Zn/Fe]_{fit}=1.1\pm0.4}$, resulting in a dust-corrected metallicity of $\mathrm{[M/H]_{tot}=-0.5\pm0.4}$.

\begin{figure*}[!htbp]
    \centering
    \includegraphics[scale=0.6]{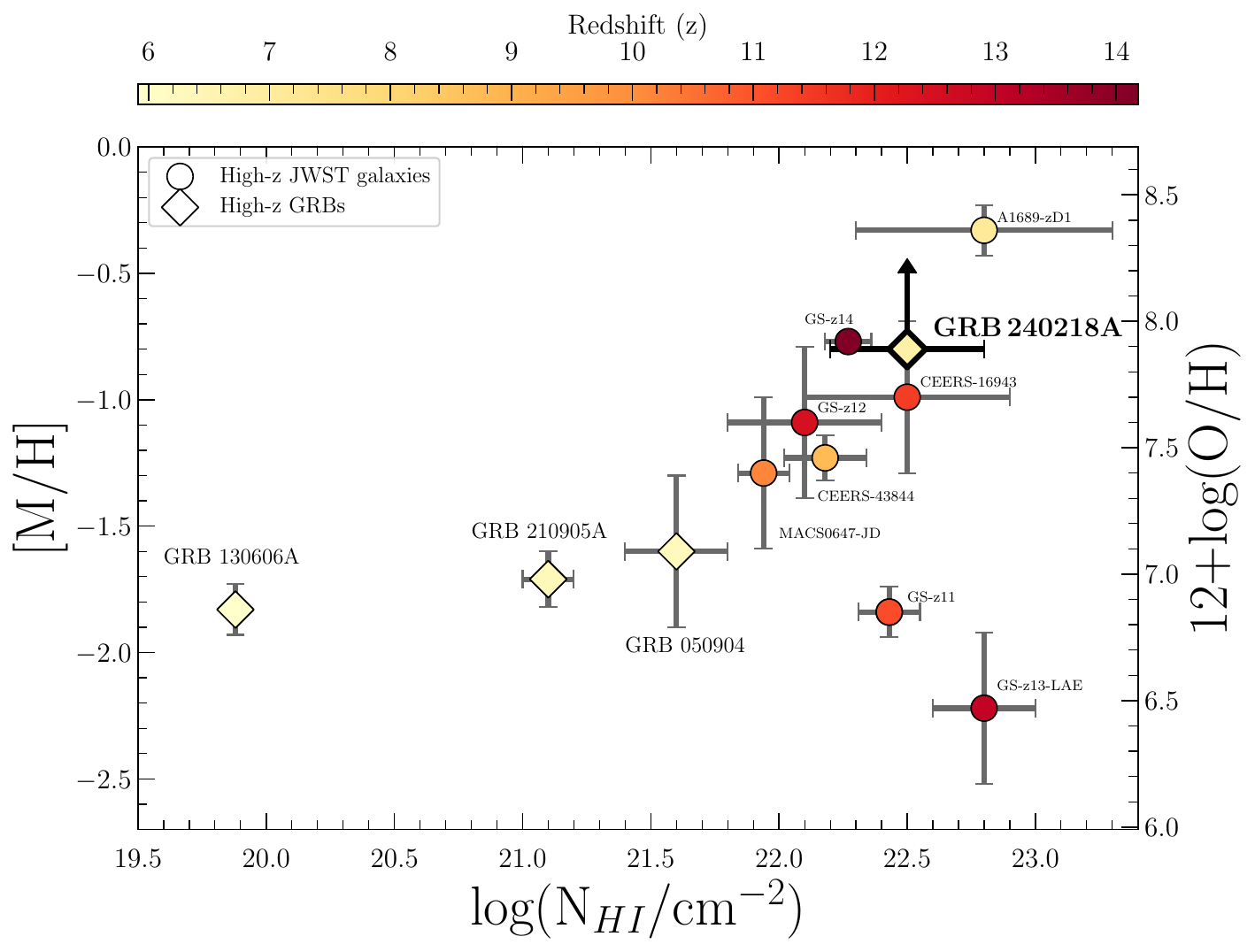}
    \caption{Metallicity  as a function of $\log (N$(\hi{})/cm$^{-2}$) color-coded by the redshift for GRBs host galaxies and high-redshift galaxies observed with JWST. For GRBs hosts, we report the observed metallicity ([M/H], not corrected for dust-depletion) determined from the afterglow spectra of GRBs at $z\gtrsim6$, i.e. GRB\,050904 (\citealt{Kawai2006,Thone2013}; note that the measurements rely on a low-resolution spectrum), GRB\,130606A (\citealt{Hartoog2015,Heintz2023}), GRB\,210905A (\citealt{Saccardi23}). For high-redshift galaxies, we report the oxygen based metallicity determined from emission lines and/or SED fitting for the recently discovered JWST high-redshift galaxies, i.e. GS-z14 (\citealt{Carniani2024_dla,Heintz2025}), GS-z12 (\citealt{D'Eugenio2024}), GS-z11 (\citealt{Hainline2024}), GS-z13-LAE (\citealt{Witstok2024_dla}), CEERS-43844, MACS0647-JD, CEERS-16943 (\citealt{Heintz2024}), A1689-zD1 (\citealt{Watson2015}; Heintz et al. in prep). The relationship between $12+$log(O/H) and [M/H] is defined by the comparison with the solar oxygen abundance, where [M/H]\,$= 12$+log(O/H)$-8.69$ \citep{Asplund2021}.}
    \label{jwst}
\end{figure*}

The host galaxy of GRB\,240218A has the highest neutral hydrogen column density measured at high-redshift \textbf{($z\gtrsim6$)} for any GRB sight-line detected thus far. The large \hi{} column density indicates that the host galaxy contains and/or is surrounded by massive, extended clouds of neutral gas, which is the raw material for star formation in galaxies. Intriguingly, such high $\log (N$(\hi{})$)$ have been reported also by \citet{Heintz2024} for three DLAs at $z>8$ observed with JWST (see Figure \ref{jwst}), and are now also detected in very high-redshift galaxies \citep[e.g.][]{Hainline2024,Witstok2024_dla}.

\begin{figure}[!htbp]
    \centering
    \includegraphics[scale=0.24]{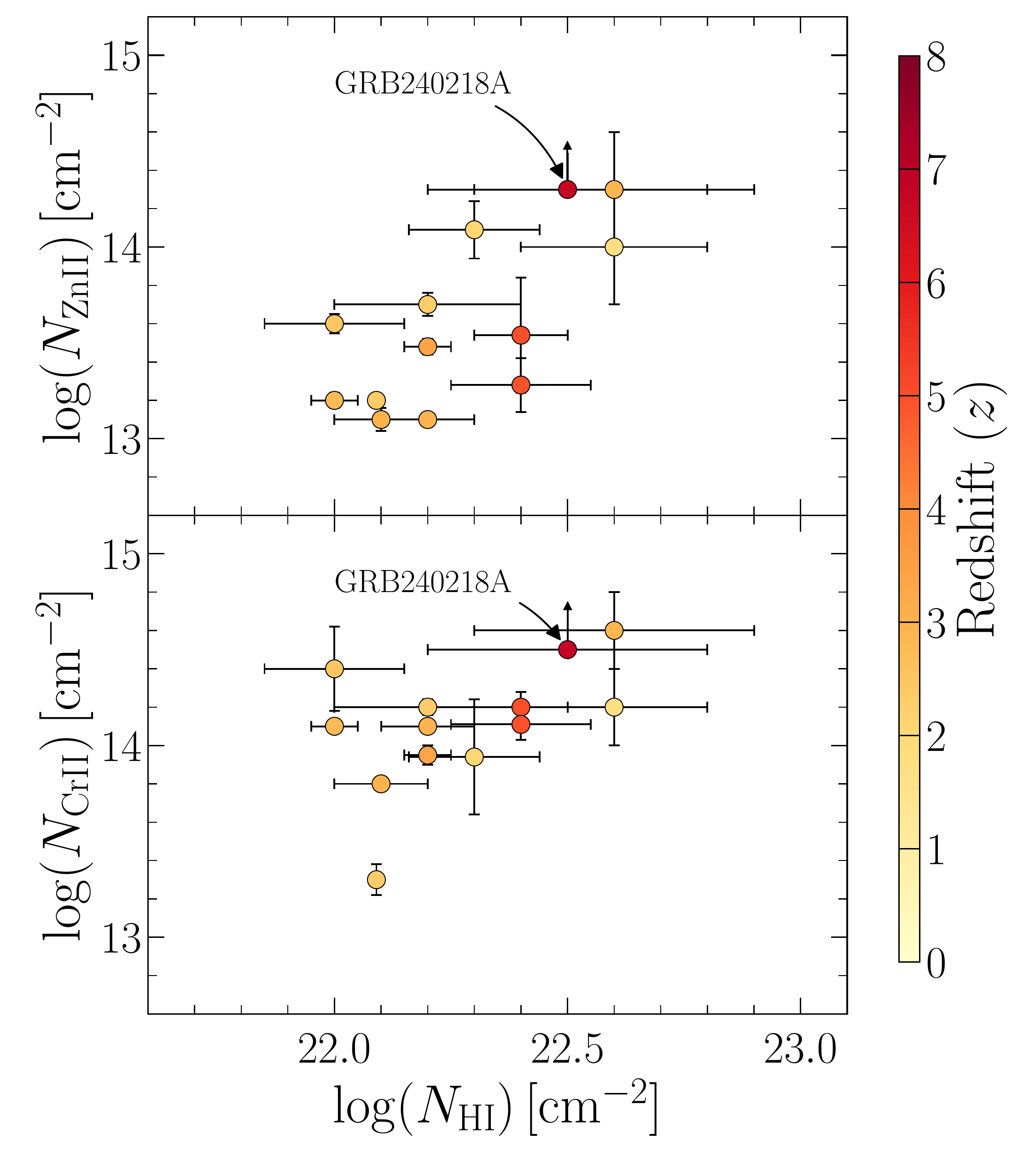}
    \caption{Column density of \znii{} and \crii{} \citep{Watson2006a,Wiseman2017,Heintz2019,Bolmer2019,Selsing2019} versus $\log (N$(\hi{})/cm$^{-2}$) \citep{Tanvir2019}, color-coded by the GRB host galaxy redshift. GRB\,240218A is marked and compared with the sample of GRBs for which these measurements are available.}
    \label{zncr}
\end{figure}

Furthermore, the large amount of metals (and likely dust) already formed at high-redshift is surprising compared to previous studies or high-$z$ simulation \citep{Heintz2023,Li2019}. This is the first time that strong absorption lines of metals, such as those of \znii{}, have been observed at high redshift. 
We investigated the extent to which the high column densities measured for \znii{} and \crii{} in the ISM of this host galaxy is likely due to the large amounts of dense gas, as suggested by the large amount of neutral hydrogen. Hence, we compared our column density measurements with those of other GRBs for which these quantities are available. We first selected GRBs for which $\log (N$(\hi{})/cm$^{-2})>22$ has been measured in \cite{Tanvir2019}, 
supplemented by more recent results. Within the selected GRBs, we then looked for the ones that had Zn and Cr column density measurements available. This led to the construction of the sample \footnote{GRB\,050401; GRB\,111008A; GRB\,120119A; GRB\,120327A; GRB\,120716A; GRB\,120815A; GRB\,140311A; GRB\,141109A; GRB\,151021A; GRB\,180325A; GRB\,181020A; GRB\,190114A} reported in Figure \ref{zncr}.
Certainly, as already stressed, the column density of \znii{}, but also of \crii{}, is an exception at high redshift. Furthermore, it also turns out to be an exception in comparison with low-redshift analogs. Indeed, from Figure \ref{zncr} it is clear that these large column densities are less common at low-redshift.

In summary, GRB\,240218A host galaxy is very likely a massive high-redshift galaxy, characterized by a large amounts of neutral gas, neutral hydrogen column and metals showing substantial, rapid chemical evolution and enrichment, as indicated by the significant presence of metals (and likely dust). Such properties could imply the need for sustained star formation before the GRB to build up the metals. The GRB is thus likely hosted by a massive galaxy that formed early. 
This GRB host galaxy is very extreme among GRB- and QSO-DLAs, which trace a common population of (low-mass) galaxies at intermediate redshifts. On the other hand, this substantial enrichment of metals and dust of the ISM is in agreement with recent findings (see Figure \ref{jwst}) based on strong nebular emission lines of star-forming regions of high-redshift galaxies at $z\sim7$ observed with JWST \citep{Langeroodi2023,Heintz2023Nat_jwst,Nakajima2023,Curti2024,Roberts-Borsani2024,Heintz2025}, as well as with the metal and dust rich high-$z$ quasars host galaxies \citep{Tripodi2023,Tripodi2024,Salvestrini2024}. The fact that such properties are found in GRB host galaxies is particularly interesting because, at high-redshift, these galaxies are selected based only on being star-forming.

While JWST usually allows the study of the metallicity of the ionized gas,
the explosion of GRB\,240218A has allowed a detailed chemical analysis of the neutral ISM of a high-redshift galaxy.
These are unprecedented measurements for a galaxy at $z\sim7$. Another advantage of GRB host galaxy studies is that, in addition to the neutral gas absorption feature, it is possible to observe the continuum and emission lines of the host galaxy, once the GRB afterglow faded. Photometric and spectroscopic observations of the host galaxy of GRB\,240218A with ALMA and JWST would provide a unique opportunity to integrate the chemical properties and the kinematics of the neutral 
gas studied in this work with those of the ionized gas \citep[e.g.][]{Schady2024}, and add unprecedented details on the chemical and physical properties of high-redshift galaxies. 
In this sense, it will also be possible to compare these future results with other GRB host galaxies at high-redshift by starting to construct a useful sample.

Our results emphasize that ground-based spectroscopic observations of absorption lines in the spectra of bright background sources, such as GRBs, are powerful tools to explore the chemical properties of the neutral gas in distant galaxies. It has been shown \citep[e.g.][]{Ghirlanda2015} that, to efficiently detect the prompt emission from high-redshift GRBs (that is the necessary first step to trigger the afterglow observations), it is essential to extend the energy range of space satellite detectors of the prompt emission to lower energies (e.g. SVOM, launched in June 2024, \citealt{Wei2016}; {\it{Einstein Probe}}, \citealt{EP_satellite,EP_mission}), and to improve sensitivity to lower flux detection limits (e.g. THESEUS, selected for ESA M7 Phase-A, \citealt{Amati2018,Tanvir2021}). We stress also that the localization capabilities of mission with such characteristics is fundamental to enhance the possibility of optical/NIR spectroscopic observations with high SNR of GRBs afterglows.  Furthermore, high-resolution absorption spectroscopy of the X-ray afterglow by the ESA large mission {\it NewAthena} can extend our knowledge of the metal abundances much closer to the GRB, thus providing a powerful diagnostics of the GRB progenitors in the high-redshift Universe \citep{Piro2022}. 

The results of our study of GRB\,240218A motivate pursuing GRB-dedicated space missions to fully exploit the possibility to use GRBs to explore the high-redshift Universe, in synergy with forthcoming ground-based instruments like the Son Of X-Shooter (SOXS, \citealt{Soxs}), and new facilities such as the future European Southern Observatory (ESO)/Extremely Large Telescope (ELT) equipped with the ArmazoNes high Dispersion Echelle Spectrograph (ANDES) high-resolution spectrograph \citep{Marconi2024_andes,D'Odorico2024_andes}. Furthermore, a novel approach to the discovery of high-redshift GRBs such as the High-z Universe GRB Observatory (HUGO, \citealt{Campana2022}) will complement the above efforts, taking advantage of the deep monitoring of the sky by the Vera Rubin Observatory \citep{Ivezic2019_llst}, to simultaneously observe exactly the same fields with a dedicated NIR facility.


\begin{acknowledgements}
A.S. acknowledges support by a postdoctoral fellowship from the CNES. S.D.V. acknowledges the support of the French Agence Nationale de la Recherche (ANR), under grant ANR-23-CE31-0011 (project PEGaSUS). L.I. acknowledges financial support from the INAF Data Grant Program ``YES'' (PI: Izzo) Multi-wavelength and multi messenger analysis of relativistic supernovae. This work has received funding from the Swiss State Secretariat for Education, Research and Innovation (SERI) under contract number MB22.00072. The Cosmic Dawn Center (DAWN) is funded by the Danish National Research Foundation under grant DNRF140. NRT acknowledges support from STFC Consolidated Grant ST/W000857/1. R.B. acknowledges funding from the Italian Space Agency, contract ASI/INAF n. I/004/11/6.
\end{acknowledgements}


\bibliographystyle{aa} 
\bibliography{mabiblio.bib} 

\appendix


\onecolumn
\section{UV pumping and photoionization model}
\label{app_a}

Absorption lines arising from metastable levels were already detected in other GRB afterglow spectra, e.g. \feii{}{*}, \niii{}{*} \citep{Vreeswijk2007,DElia2009,Delia2010,DElia2011,DElia2014}. As fine-structure lines, the population of the corresponding levels has been shown to be due to the UV radiation \citep{Vreeswijk2007,DElia2009}. Together with fine-structure lines, excited absorption features produced by metastable levels are used to determine the distance of the closest gas clouds from the GRB responsible for the UV/optical absorption lines \citep{Vreeswijk2007,Prochaska2008,DElia2009,Vreeswijk2012,Vreeswijk2013}.
Since the lifetime of high quantum levels state is short, during the spontaneous decay process lower levels are populated, either with principal quantum number $n>1$ (excited states), or with $n=1$ but high values of the spin-orbit coupling (the so-called fine-structure levels of the ground state). 
This process is known as indirect UV pumping and is opposed to the other mechanism able to populate higher levels of the ions and atoms: the collisional processes. 
The evidence of UV pumping as the primary factor responsible for the population of fine-structure/metastable levels 
this excitation comes from the variability of the fine-structure levels, which 
has been reported every time multi-epoch spectroscopy was available \citep[e.g.][]{Prochaska2006, Vreeswijk2007, DElia2009_variab}. 

For GRB\,240218A 
we have (see Table \ref{table_N}) the detection of the \feii{} ground state, the first fine-structure level (marked as 1s), and the ground state of the $n=2$ first excited state (marked as 5s). These levels have been detected both for component II and III. Instead, no absorption lines are detected at the wavelength corresponding to the ground state of the metastable transitions of \niii{}. The low SNR and the strong atmospheric absorption in the region of the spectrum where the \niii{} multiplets (\niii{}\,$\lambda1317$, \niii{}\,$\lambda1370$, \niii{}\,$\lambda1454$ and \niii{}\,$\lambda1709$, \niii{}\,$\lambda1741$, \niii{}\,$\lambda1751$) fall, prevent us from determining how much \niii{} is in the ground state compared to the excited levels. The scenario in which the \niii{} ground state is absent and the \niii{} is all in the excited states could be explained by the fact that it takes a long time for the ion to decay to the \niii{} ground state, $\sim 37$\,h \citep{Vreeswijk2007}. 
Hence, transitions from \niii{}, and also \feii{}, metastable levels serve as strong indicators of the UV pumping mechanism, as they can be detected several hours post-GRB event.
Unfortunately, we do not have multi-epoch spectroscopy. However, we note that the \feii{} first excited state (5s) has a column density higher than that of the fine structure levels (Table \ref{table_N}). Since collisional processes populate levels according to a Boltzmann distribution, with the lower energy levels being the most populated, this inversion of the population can be interpreted as another signature of the UV pumping mechanism.

The input of the code is the initial column density of the \feii, which is assumed to be completely in the ground state before it is hit by the GRB afterglow and the Doppler parameter of the absorbing gas. These data have been taken from Table \ref{table_N} (for the initial column density we summed the columns of the three \feii\,levels). We stress that the estimated distances derived from the photoionization model could be larger if the possible saturation issue is not negligible. 

\section{Curve of growth analysis}
\label{app_b}

\begin{figure}[!htbp]
    \centering
    \includegraphics[scale=0.3]{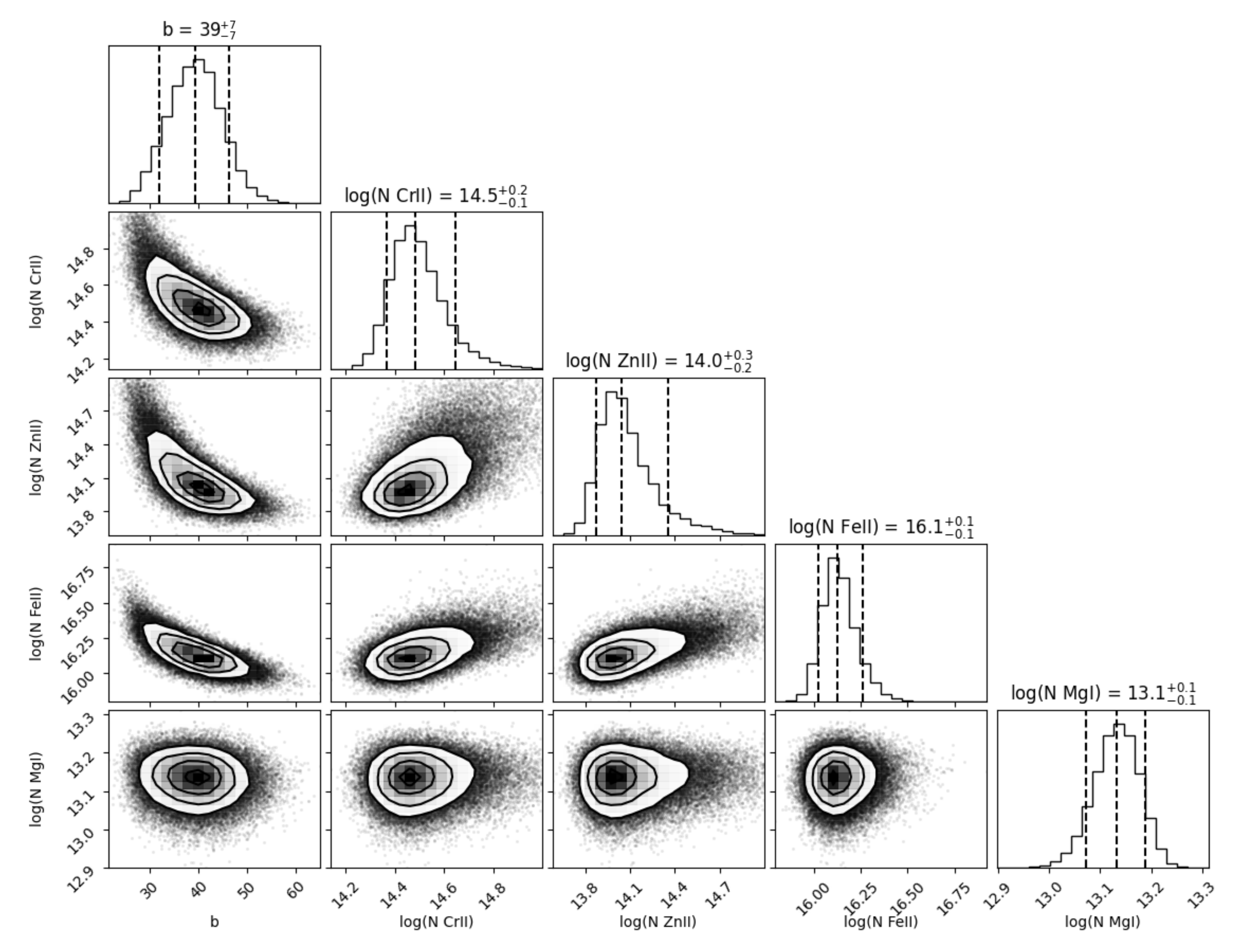}
    \caption{Corner plot showing the results of our MCMC parameter estimation. The histograms show the marginalized posterior densities for each parameter, while the two-dimensional projection reveal covariances.
    The best-fit parameters are  indicated with black dashed lines (50th percentile). The uncertainties, shown also with black dashed lines, are computed as the 16th and 84th percentiles of the posterior samples along each axis, thus representing 1$\sigma$ confidence ranges.}
    \label{mcmc_cog}
\end{figure}

The CoG (for an exhaustive description of the method please refer to eq. 3-50, 3-51 of \citealt{Spitzer1998}) can be segmented into three smoothly connected regimes, each illustrating a distinct relationship between $EW$ and $N$: at low $EW$ values ($< 0.1 \AA$), $EW$ is linearly correlated with column density. For intermediate values, the correlation becomes logarithmic ($EW \propto \log(\tau) \propto N_i / b$), and at high values, we observe $EW \propto \sqrt{\tau}$. In the latter two regimes, absorption lines are affected by saturation, making precise $N_i$ estimation challenging, as small $EW$ variations correspond to large $N_i$ changes. This issue can be mitigated by measuring multiple lines from the same transition, particularly if they originate from the ground state to the first excited level, where ions have similar excitation energies and share a common kinetic distribution. A limitation of this method, however, is the lack of detailed velocity distribution information for the multiple ISM clouds contributing to the observed absorption lines.
In the spectrum of GRB\,240218A, we identified three sets of spectral transitions, \feii, \znii, and \crii, originating from the ground state to the first excited state. These transitions have consistent column densities and ionization potentials. We measured the equivalent widths of these absorption lines (presented in Table \ref{tab:EW}), noting that \znii\,$\lambda2062$ and \crii\,$\lambda2062$ are blended. To estimate their individual contributions, we first measured the total blend width ($EW = 7.68 \AA$) and then apportioned contributions based on the oscillator strength ratios for \znii\,and \crii. The same procedure has been adopted for \znii\,$\lambda2026$\,and \mgi\,$\lambda2026$.
Finally, we fit the CoG with free parameters for $b$ and the column densities of Fe, Zn, and Cr using a uniform prior within a Monte Carlo Markov Chain (MCMC) sampling framework implemented in Python {\em emcee} package \citep{emcee}. For further details on the computational method, refer to \citet{dewet2023}. The results from the CoG supports and confirms the results obtained using Voigt Fit of absorption lines. The posterior distributions of all parameters derived from the MCMC procedure are shown in Figure \ref{mcmc_cog}.

\begin{table}[!htbp]
\centering
\caption{Rest frame Equivalent Widths of the identified absorption lines used to perform the curve of growth analysis.}
\begin{tabular}{|c|c|c|c|}
\hline
Element & Transition & Observed Wavelength [\AA] & Equivalent Width ($EW_r$) [\AA] \\
\hline
\feii & 1608 & 12517.22 & 0.91 $\pm$ 0.09 \\
\feii & 1611 & 12538.63 & 0.32 $\pm$ 0.03 \\
\feii & 2260 & 17593.76 & 0.62 $\pm$ 0.07 \\
\znii & 2026 & 15768.40 & 0.90 $\pm$ 0.08 \\
\znii & 2062 & 16048.32 & 0.50 $\pm$ 0.09 \\
\crii & 2056 & 16001.72 & 0.58 $\pm$ 0.06 \\
\crii & 2062 & 16051.65 & 0.49 $\pm$ 0.08 \\
\mgi & 2026 & 15770.37 & 0.05 $\pm$ 0.01 \\
\hline
\end{tabular}
\label{tab:EW}
\end{table}

\section{Nucleosynthesis and component-by-component analysis}

\label{app_d}

\begin{figure}[h!]
    \centering
    \includegraphics[scale=0.6]{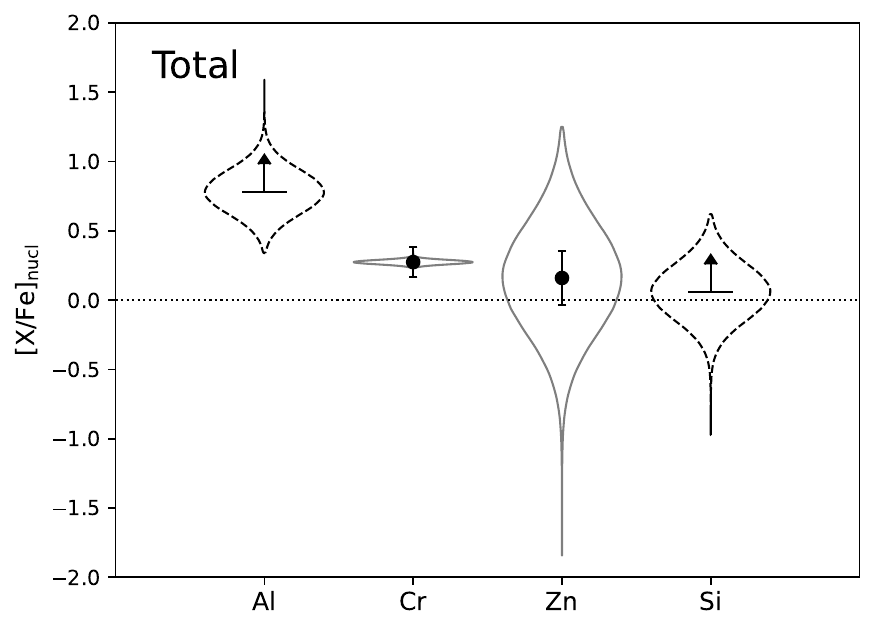}
    \caption{Abundances of different elements with respect to iron after correcting for dust depletion, for the total analysis. Limits are indicated by arrows and the error bars represent the uncertainty propagated from column density measurements; the violin plots represents the uncertainty caused by dust depletion estimated using one million Monte Carlo realizations of the $\mathrm{[Zn/Fe]_{fit}}$.}
    \label{total_violin}
\end{figure}

We inspect also the [X/Fe] residuals (over- and under-abundance of different elements with respect to iron i.e. after correcting for dust depletion), as shown in Figure \ref{total_violin}. They represents the deviations from the linear fits of Figure \ref{dust_240218A_measure} which are likely due to the effects of nucleosynthesis, or peculiar abundances in the host ISM. The corresponding values and errors are reported in Table \ref{table_abb_depl}. Given the small number of elements available we do not compare our results with nucleosynthetic models.

\begin{figure}[h!]
    \centering
    \includegraphics[scale=0.4]{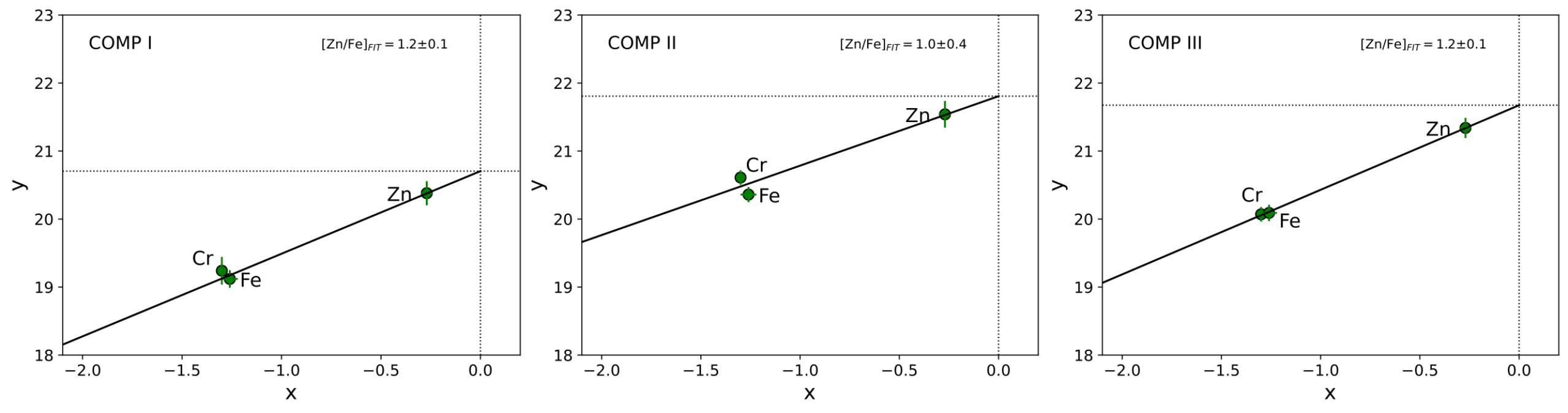}
    \caption{Similar to Figure \ref{dust_240218A_measure}, but for the individual components of the absorption-line profile. In this case $y$ can be interpreted as an equivalent metal column (given that the information on the \hi{} is not available)) and it is defined in detail in \cite{Ramburuth-Hurt2023}. Again, the slope of the linear fit to the data (solid line) determines the overall strength of depletion $\mathrm{[Zn/Fe]_{fit}}$, as labeled.}
    \label{fit_comp}
\end{figure}

\begin{figure}[h!]
    \centering
    \includegraphics[scale=0.4]{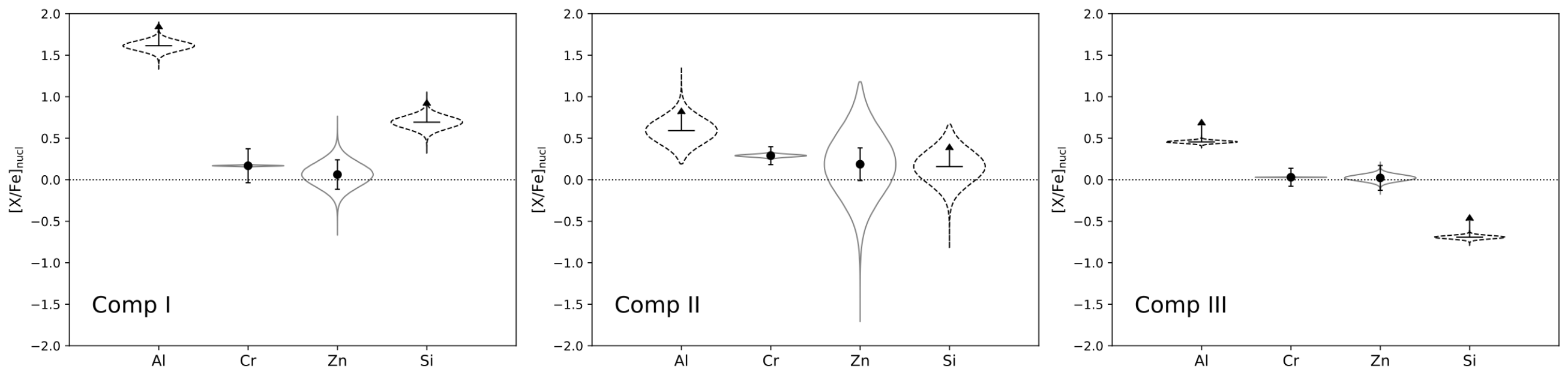}
    \caption{Same as Figure \ref{total_violin} but for component-by-component analysis.}
    \label{abund_comp}
\end{figure}

\begin{table*}
\caption{[X/Fe] residuals of the depletion pattern fitting (see Figures \ref{dust_240218A_measure} and \ref{fit_comp}). The values and uncertainties are obtained including MC simulations to take into account the dust depletion errors.}
\label{table_abb_depl} 
\centering
\begin{tabular}{c c c c c}       
\hline\hline
X  & $I$ & $II$ & $III$ & Tot\\
\hline
\rule{0pt}{3ex}
[Al/Fe]  &   $1.6^{+0.1}_{-0.1}$  & $0.6^{+0.2}_{-0.2}$ & $0.5^{+0.1}_{-0.1}$ &  $0.8^{+0.2}_{-0.2}$\\[2ex]

[Cr/Fe]  &   $0.2^{+0.1}_{-0.1}$  & $0.3^{+0.1}_{-0.1}$ & $0.0^{+0.1}_{-0.1}$  &  $0.3^{+0.1}_{-0.1}$\\[2ex]

[Zn/Fe] &   $0.1^{+0.2}_{-0.2}$   & $0.2^{+0.4}_{-0.4}$ & $ 0.0^{+0.1}_{-0.1}$  &  $0.2^{+0.4}_{-0.4}$\\[2ex]

[Si/Fe] &   $0.7^{+0.1}_{-0.1}$  & $0.2^{+0.2}_{-0.2}$  & $-0.7^{+0.1}_{-0.1}$  &  $0.1^{+0.2}_{-0.2}$\\[2ex]

\hline
\hline
\end{tabular}
\end{table*}

\end{document}